\documentclass[showpacs,showkeys,twocolumn]{revtex4}

\usepackage{amsfonts,amsmath,amssymb,bm,hyperref}

\begin{document}


\title{Evolution of coupled classical fields}

\author{Maxim Dvornikov}
\altaffiliation{Fachbereich Physik, Universit\"{a}t Rostock, 18051
Rostock, Germany} \email{maxim.dvornikov@uni-rostock.de}

\date{\today}

\begin{abstract}
We study the evolution of the coupled scalar and fermion fields
within the classical field theory. We examine the case of $N$
coupled fields in $1+3$ dimensional space. The general expressions
for the fields distributions are obtained. The particular case of
two fields in $1+1$ dimensional space is carefully studied. We
obtain the expressions for the averaged fields intensities and
show that in the relativistic limit they are similar to the usual
transition probabilities formulae of neutrino oscillations.
\end{abstract}



\maketitle

The particle mixing plays an important role in elementary particle
physics. According to the experimental data the particle mixing
exists in both quark and lepton sectors of the standard model. The
idea of mixing among the two quark flavors was put forward in
Ref.~\cite{Cab63} to explain the baryons decays. The mixing in the
leptonic sector of the standard model was proposed in
Ref.~\cite{Pon58eng}. In that parer the neutrino mixing and
oscillations were studied on the analogy of the known at that time
$K^0\leftrightarrow\bar{K}^0$ oscillations. Then in
Ref.~\cite{MakNakSak62} this approach was generalized on the
mixing between three neutrino flavors. Recently we obtained the
strong evidence in favor of neutrino oscillations and, therefore,
mixing (see, for instance, Ref.~\cite{Ahm04}). For example,
neutrino oscillations are likely to be the most plausible
explanation of the solar and atmospheric neutrino problems.

In Ref.~\cite{Pon58eng} the neutrino oscillations were examined
within the quantum mechanical approach. Schr\"{o}dinger like
differential equation for the description of the two-level
neutrino system was proposed. On the basis of this equation one
can derive the famous transition probability formula,
\begin{equation}\label{TrProbPon}
  P(t)=\sin^2(2\theta_\mathrm{vac})
  \sin
  \left(
    \frac{\Delta m^2}{4E}t
  \right),
\end{equation}
where $\theta_\mathrm{vac}$ is the vacuum mixing angle, $\Delta
m^2$ is the mass squared difference and $E$ is the energy of the
system. Up to now Eq.~\eqref{TrProbPon} is of use in numerous
phenomenological studies of neutrino oscillations. However more
profound analysis of particle mixing and oscillations is
necessary. The approach to the oscillations phenomenon based on
the field theory methods should be elaborated.

In the last decade a great deal of studies on the field
theoretical substantiation of Eq.~\eqref{TrProbPon} were carried
out. First of all it is necessary to mention works by M.~Blasone
and G.~Vitiello and their collaborators (see
Refs.~\cite{BlaVit95,BlaHenVit99,BlaVit99,BlaPacTse03}). In these
papers the authors made the comprehensive analysis of the fermion
and boson mixing transformations using the methods of quantum
field theory. It was demonstrated that the vacuum structure of
mass eigenstates is not equivalent to one of flavor eigenstates.
The quantum mechanical formula for the transition probability was
also reproduced. Moreover some corrections, which result from more
careful quantum field theory analysis, were obtained. In
Ref.~\cite{BlaCapVit02} the group theoretical aspects of neutrino
oscillations were discussed. Analogous approach to description of
the neutrino oscillations was developed in
Refs.~\cite{FujHabYab99,FujHabYab01}.

Rather appreciable contributions to the investigation of the
flavor neutrino oscillations were made in
Refs~\cite{GiuKim98,Giu03}. In those papers the wave packages
treatment of neutrino oscillations was developed as well as the
discussion of the neutrino oscillations phase is presented. The
neutrino oscillations phase was also studied in
Ref.~\cite{OkuSchTsu03}. A very interesting approach to the
description of the neutrino flavor oscillations was proposed in
Ref.~\cite{Fie03EPJ}. In that paper the covariant path amplitudes
method was applied for the analysis of the neutrino oscillations
phase.

Recently we elaborated the quasi-classical approach for the
description of spin (see
Refs.~\cite{EgoLobStu00,LobStu01,DvoStu02JHEP}) and flavor
neutrino oscillations (see Ref.~\cite{GriLobStu01}). It was shown
that neutrino oscillations in moving and polarized matter under
the influence of arbitrary electromagnetic fields were described
by the generalized Lorentz invariant quasi-classical
Bargmann-Michel-Telegdi equation. It is interesting to note that
the equation describing the precession of the neutrino three
dimensional spin vector (neutrino spin in particle's rest frame)
is the usual Bloch equation. We demonstrated that neutrino spin
rotates around a certain direction determined by the velocities
and polarizations of background fermions as well as the
electromagnetic field strength.

The method involving the Bloch equation for the treatment of
neutrino flavor oscillations was proposed in Ref.~\cite{Sto87}. If
one considers the evolution of two neutrinos system (e.g., $\nu_e$
and $\nu_\mu$), it is possible to introduce the "polarization"
vector $\mathbf{P}=\mathrm{Tr}(\bm{\sigma}\rho)$, where
$\bm{\sigma}$ are the Pauli matrices and $\rho$ is the $2\times 2$
density matrix. If neutrinos propagate in vacuum, the vector
$\mathbf{P}$ was shown to precess without loss of length according
to the Bloch equation. The appearance of classical effects in
various quantum systems (including the analysis of a two level
system with help of the Bloch equation) was discussed in
Ref.~\cite{Giu96}. Thus basing on the similarity in the
description of neutrino spin and flavor oscillations we suppose
that classical theory methods could have been applied for the
treatment of the flavor oscillations. However this supposition
should be substantiated by the direct calculations that show the
classical theory yields at least the same results as the quantum
one.

In this paper we study the evolution of the coupled scalar as well
as fermion fields within the context of classical field theory.
The main goal of our article is to demonstrate that neutrino
oscillations can be described within the classical approach. The
classical approach was also adopted since we should not be puzzled
by a problem: must we rely on flavor or mass eigenstates in our
treatment of neutrino oscillations? The intensive discussion about
this topic takes place in Refs.~\cite{Giu03,Fie03}. The case of
$N$ coupled fields in $1+3$ dimensional space is examined. We
solve the Cauchy problem for this system, i.e. for the given
initial conditions we find the fields distributions for any time
point. In order to analyze the obtained expressions we study the
particular case of two fields in $1+1$ dimensional space. For the
specific initial conditions the expressions for the averaged
fields intensities are obtained. We also show that in the
relativistic limit they are similar to the usual transition
probabilities formulae of neutrino oscillations in vacuum. It is
interesting to mention that the expressions for the averaged
fields intensities are identical for both bosons and fermions.

First let us discuss the case of $N$ arbitrary coupled scalar
fields. For simplicity we suppose that these fields are the real
ones. The Lagrangian for this system is expressed in the following
form
\begin{equation}\label{LagrphiN}
  \mathcal{L}(\bm{\varphi})=
  \sum_{k=1}^{N}\mathcal{L}_0(\varphi_k)+
  \sum_{\substack{i,j=1 \\ i\neq k}}^{N}
  g_{ik}\varphi_i\varphi_k,
\end{equation}
where $g_{ik}$ are the coupling constants,
$\bm{\varphi}=(\varphi_1,\dots,\varphi_N)$, and
\begin{equation}\label{Lagr0phi}
  \mathcal{L}_0(\varphi_k)=
  \frac{1}{2}\partial_\mu\varphi_k\partial^\mu\varphi_k-
  \frac{\mathfrak{m}_k^2}{2}\varphi_k^2,
\end{equation}
is the Lagrangian for the field $\varphi_k(\mathbf{r},t)$ at the
absence of the additional coupling, $\mathfrak{m}_k$ is the mass
corresponding to this field. It is necessary to note that the
second term in Eq.~\eqref{LagrphiN} is assumed to be an
interaction between fields $\varphi_k$.

In order to describe the evolution of the system
\eqref{LagrphiN}-\eqref{Lagr0phi} we should set the Cauchy problem
for this system. For the initial conditions,
\begin{equation}\label{inicond}
  \varphi_i(\mathbf{r},0)=f_i(\mathbf{r}),
  \quad
  \dot{\varphi}_i(\mathbf{r},0)=g_i(\mathbf{r}),
\end{equation}
where $f_i(\mathbf{r})$ and $g_i(\mathbf{r})$ are the given
functions, one should find the fields distributions
$\varphi_k(\mathbf{r},t)$ for any time point.

It is always possible to diagonalize the Lagrangian
\eqref{LagrphiN} with help of the transformation,
\begin{equation*}
  \varphi_i(\mathbf{r},t)=
  \sum_{k=1}^{N}M_{ik}u_k(\mathbf{r},t).
\end{equation*}
Thus the Lagrangian expressed in terms of the fields
$u_k(\mathbf{r},t)$ takes the form
\begin{equation*}
  \mathcal{L}(\mathbf{u})=
  \sum_{k=1}^{N}\mathcal{L}_0(u_k),
\end{equation*}
where $\mathcal{L}_0(u_k)$ is the Lagrangian for the field
$u_k(\mathbf{r},t)$,
\begin{equation*}
  \mathcal{L}_0(u_k)=
  \frac{1}{2}\partial_\mu u_k\partial^\mu u_k-
  \frac{m_k^2}{2}u_k^2,
\end{equation*}
and $m_k$ are the corresponding masses. It should be noted that
these masses differ from the masses of the fields $\varphi_k$. The
fields $u_k(\mathbf{r},t)$ are usually called mass eigenstates in
contrast to $\varphi_k(\mathbf{r},t)$.

One can write the differential equations for the fields
$u_k(\mathbf{r},t)$. It is the system of the usual homogeneous
Klein-Gordon equations. Their solutions have the form
\begin{equation}\label{usol}
  u_k(\mathbf{r},t)=
  \int
  \frac{\mathrm{d}^3\mathbf{p}}{(2\pi)^3}
  \left[
    a_k^{+}(\mathbf{p})e^{-i\mathcal{E}_k t}+
    a_k^{-}(\mathbf{p})e^{i\mathcal{E}_k t}
  \right]
  e^{i\mathbf{p}\mathbf{r}}
\end{equation}
where $\mathcal{E}_k=\sqrt{\mathbf{p}^2+m_k^2}$, and
$a_k^{\pm}(\mathbf{p})$ are the Fourier coefficients. Note that
$a_k^{\pm}(\mathbf{p})$ are the \textit{c}-numbers.

To solve the Cauchy problem we introduce the functions
\begin{align*}
  F_k(\mathbf{r}) &=
  \sum_{i=1}^N
  \left(
  M^{-1}
  \right)_{ki}
  f_i(\mathbf{r}),
  \\
  G_k(\mathbf{r}) &=
  \sum_{i=1}^N
  \left(
  M^{-1}
  \right)_{ki}
  g_i(\mathbf{r}).
\end{align*}
These functions are the initial conditions for the fields $u_k$.
Then one should pick out the coefficients $a_k^{\pm}(\mathbf{p})$
so that to satisfy the initial conditions \eqref{inicond}. From
Eqs.~\eqref{inicond} and \eqref{usol} we obtain
\begin{equation*}
  a_k^{\pm}(\mathbf{p})=
  \frac{1}{2}
  \left(
  F_k(\mathbf{p})\pm i
  \frac{G_k(\mathbf{p})}{\mathcal{E}_k}
  \right),
\end{equation*}
where $F_k(\mathbf{p})$ and $G_k(\mathbf{p})$ are the Fourier
transforms of the functions $F_k(\mathbf{r})$ and
$G_k(\mathbf{r})$ respectively. Finally we receive the fields
distributions $\varphi_j(\mathbf{r},t)$ in the explicit form
\begin{align}
  \varphi_j(\mathbf{r},t)= &
  \sum_{ik=1}^N
  M_{jk}
  \left(
  M^{-1}
  \right)_{ki}
  \notag
  \\
  &
  \times
  \int \mathrm{d}^3\mathbf{r}'
  \big[
  \dot{D}_k(\mathbf{r}-\mathbf{r}',t)f_i(\mathbf{r}')
  \notag
  \\
  \label{phisol3D}
  & +
  D_k(\mathbf{r}-\mathbf{r}',t)g_i(\mathbf{r}')
  \big],
\end{align}
where
\begin{equation}\label{PJfunc4D}
  D_k(\mathbf{r},t)=
  \int
  \frac{\mathrm{d}^3\mathbf{p}}{(2\pi)^3}
  e^{i\mathbf{p}\mathbf{r}}
  \frac{\sin\mathcal{E}_k t}{\mathcal{E}_k},
\end{equation}
is the Pauli-Jordan function. It is interesting to list some of
the properties of the Pauli-Jordan function
\begin{equation*}
  D_k(\mathbf{r},0)=0,
  \quad
  \dot{D}_k(\mathbf{r},0)=\delta^3(\mathbf{r}),
  \quad
  \ddot{D}_k(\mathbf{r},0)=0.
\end{equation*}
It is worth noticing that the initial conditions in
Eq.~\eqref{inicond} are consistent with these properties of the
Pauli-Jordan function. We also mention that the Pauli-Jordan
function can be expressed in the explicit form (see, e.g.,
Ref.~\cite{BogShi80p603}),
\begin{equation}\label{PJfunEF}
  D_k(\mathbf{r},t)=
  \frac{1}{2\pi}
  \varepsilon(t)\delta(s^2)-
  \frac{m_k}{4\pi s}
  \varepsilon(t)\theta(s^2)J_1(m_k s),
\end{equation}
where $s^2=t^2-\mathbf{r}^2$ and
\begin{equation*}
  \varepsilon(t)=
  \begin{cases}
    1, & t>0, \\
    -1, & t<0.
  \end{cases}
  \qquad
  \theta(s)=
  \begin{cases}
    1, & s>0, \\
    0, & s<0.
  \end{cases}
\end{equation*}
are the step functions. Thus Eqs.~\eqref{phisol3D} and
\eqref{PJfunc4D} represent the exact solution of the Cauchy
problem for arbitrary functions $f_i(\mathbf{r})$ and
$g_i(\mathbf{r})$.

The integrals calculation in Eq.~\eqref{phisol3D}, however, are
rather awkward in general $1+3$ dimensional space. Thus let us,
for simplicity, consider the space with $1+1$ dimensions. Instead
of Eq.~\eqref{PJfunc4D} we have
\begin{equation*}
  D_k(x,t)=
  \int_{-\infty}^{+\infty}
  \frac{\mathrm{d}p}{2\pi}
  e^{ipx}
  \frac{\sin\mathcal{E}_k t}{\mathcal{E}_k}.
\end{equation*}
Now $\mathcal{E}_k=\sqrt{p^2+m_k^2}$. One can also obtain the
Pauli-Jordan function in the explicit form [see
Eq.~\eqref{PJfunEF}] in $1+1$ dimensional space,
\begin{align}
  \label{Dk}
  D_k(x,t) &=
  \frac{1}{2}
  \theta(s^2)J_0(m_k s),
  \\
  \label{Dkdot}
  \dot{D}_k(x,t) &=
  t\delta(s^2)-
  \frac{m_k t}{2s}
  \theta(s^2)J_1(m_k s).
\end{align}
Here $s^2=t^2-x^2$.

We suppose that $g_k(x)=0$ and $f_k(x)\neq 0$. Then we receive
(see also Ref.~\cite{MorFes53p854})
\begin{align}
  \varphi_j(x,t)= &
  \sum_{ik=1}^N
  M_{jk}
  \left(
  M^{-1}
  \right)_{ki}
  \notag
  \\
  &
  \times
  \Big\{
  \frac{1}{2}
  \left[
  f_i(x-t)+f_i(x+t)
  \right]
  \notag
  \\
  &
  \label{phisol1+1}
  -\frac{m_k t}{2}
  \int_{x-t}^{x+t} \mathrm{d}y
  \thinspace f_i(y)
  \frac{J_1(m_k s)}{s}
  \Big\}.
\end{align}
It should be noted that if functions $f_i(x)\not=0$ in a bounded
region and $f_i(x)\to 0$, when $x\to\pm\infty$ then the second
term in Eq.~\eqref{phisol1+1} is the vanishing one and
\begin{equation*}
  \varphi_j(x,t)\to
  \frac{1}{2}
  [f_j(x-t)+f_j(x+t)].
\end{equation*}
This property of the Klein-Gordon equation was also mentioned in
Ref.~\cite{MorFes53p854}. The described feature has one
interesting physical implication. If a single particle appears far
from a detector, then its field distribution is localized in
space. When a particle begins propagating towards a detector its
field distribution approaches to the initial conditions. Thus the
effect of various non-trivial phenomena (like conversion, or
oscillations, from one field type to another) will be vanishing.

Now let us choose the initial conditions. We suppose $f_1(x)=0$
and
\begin{equation*}
  f_2(x)=\mathfrak{A}\sin
  \left(
    \frac{\omega}{2}x
  \right),
  \quad
  \mathfrak{A}=\frac{4}{\sqrt{\omega L}},
\end{equation*}
where $L$ is the "volume" of the space. Note that $\mathfrak{A}$
is just the normalization factor. In this case we can calculate
the integral in Eq.~\eqref{phisol1+1} explicitly
\begin{multline}\label{IntermInt}
  \int_{x-t}^{x+t} \mathrm{d}y
  \thinspace \sin
  \left(
    \frac{\omega}{2}y
  \right)
  \frac{J_1(m s)}{s}
  \\
  =
  \pi\sin
  \left(
    \frac{\omega}{2}x
  \right)
  J_{1/2}
  \left(
  \frac{t}{2}
  \beta_1
  \right)
  J_{1/2}
  \left(
  \frac{t}{2}
  \beta_2
  \right),
\end{multline}
where
\begin{equation*}
  \beta_{1,2}=\sqrt{\frac{\omega^2}{4}+m^2}\mp
  \frac{\omega}{2}.
\end{equation*}
In computation of the integral in Eq.~\eqref{IntermInt} we used
the expression
\begin{multline*}
  \int_{0}^{a}
  \mathrm{d}x
  \frac{\cos(b\sqrt{a^2-x^2})}{\sqrt{a^2-x^2}}J_\nu(cx)
  \\
  =
  \frac{\pi}{2}
  J_{\nu/2}
  \left[
    \frac{a}{2}
    \left(
      \sqrt{b^2+c^2}-b
    \right)
  \right]
  J_{\nu/2}
  \left[
    \frac{a}{2}
    \left(
      \sqrt{b^2+c^2}+b
    \right)
  \right].
\end{multline*}
One half order Bessel function can be expressed in terms of the
elementary function. Namely,
\begin{equation}\label{J1/2}
  J_{1/2}(z)=
  \sqrt{\frac{2}{\pi z}}\sin z.
\end{equation}

Let us consider the case when $\omega$ has great values compared
to the masses $m_{1,2}$: $\omega\gg m_{1,2}$. This situation
corresponds to the high energy approximation or relativistic
"particles". Then, the parameters $\beta_{1,2}$ take the form
\begin{equation}\label{beta12}
  \beta_{1}=\frac{m^2}{\omega},
  \quad
  \beta_{2}=\omega.
\end{equation}

The field distribution $\varphi_1(x,0)$ is equal to zero. Thus we
can describe the dynamics of this field for the subsequent points
of time. If one studies the evolution of two fields,
$\varphi_{1,2}(x,t)$, it is possible to parameterize the matrix
$M_{jk}$ with help of one angle
\begin{equation}\label{Mtheta}
  M_{jk}=
  \begin{pmatrix}
    \cos\theta & \sin\theta \\
    -\sin\theta & \cos\theta
  \end{pmatrix}.
\end{equation}
Using Eqs.~\eqref{IntermInt}-\eqref{Mtheta} we can rewrite
Eq.~\eqref{phisol1+1} in the following way
\begin{align}
  \varphi_1(x,t)= & 2\mathfrak{A}\sin 2\theta
  \sin
  \left(
    \frac{\omega}{2} x
  \right)
  \sin
  \left(
    \frac{\omega}{2} t
  \right)
  \notag
  \\
  &
  \notag
  \times
  \sin
  \left[
  \frac{t}{4\omega}(m_1^2-m_2^2)
  \right]
  \\
  &
  \label{phinotaver}
  \times
  \cos
  \left[
  \frac{t}{4\omega}(m_1^2+m_2^2)
  \right].
\end{align}

Now let us discuss the field measurement process. In case of
rapidly varying fields ($\omega\gg m_{1,2}$), a detector registers
not the field strength, but the intensity of the field which is
proportional to the field strength squared, $I\sim\varphi^2(x,t)$.
Moreover a detector has limited sensitivity, i.e. it cannot
register arbitrary field variation in time and in space. Thus we
should average the intensity over the characteristic time and
space scales. These scales should be greater than typical time and
space scales of the field in question, i.e. $1/\omega$.

To calculate the mean value of the intensity one should take into
account the expressions
\begin{equation*}
  \left\langle
    \sin^2
    \left(
      \frac{\omega}{2} x
    \right)
  \right\rangle=
  \left\langle
    \sin^2
    \left(
      \frac{\omega}{2} t
    \right)
  \right\rangle=
  \frac{1}{2},
\end{equation*}
and
\begin{gather*}
  \left\langle
  \sin^2
  \left[
  \frac{t}{4\omega}(m_1^2-m_2^2)
  \right]
  \right\rangle=
  \sin^2
  \left[
  \frac{t}{4\omega}(m_1^2-m_2^2)
  \right],
  \\
  \left\langle
  \cos^2
  \left[
  \frac{t}{4\omega}(m_1^2+m_2^2)
  \right]
  \right\rangle=
  \cos^2
  \left[
  \frac{t}{4\omega}(m_1^2+m_2^2)
  \right],
\end{gather*}
since $(m_1^2\pm m_2^2)/\omega\ll \omega$. Then we should
introduce the normalized intensity of the field $\varphi_1(x,t)$
according to the formula
\begin{equation*}
  P(t)=\frac{\langle I \rangle (t)}{\mathfrak{A}^2}.
\end{equation*}
Finally we obtain the expression for $P(t)$ in the following form,
\begin{align}
  P(t)= &
  \sin^2 2\theta
  \sin^2
  \left(
    \frac{\Delta m^2}{4\omega}t
  \right)
  \notag
  \\
  &
  \label{Intens}
  \times
  \left\{
    1-
    \sin^2
    \left(
      \frac{m_1^2+m_2^2}{4\omega}t
    \right)
  \right\},
\end{align}
where we introduced the common notation $\Delta m^2=m_1^2-m_2^2$.

It is necessary to identify the $\omega$ parameter. We cannot
directly equate it to the particle energy, $E=\hbar\omega$, since
we are using the classical approach here. Moreover, the chosen
"wave function", $f_2(x)\sim\sin(\omega x/2)$, does not correspond
to a definite momentum and thus to a definite energy. However we
can calculate the averaged energy density of the system,
\begin{equation}\label{enerdens}
  \langle\rho_E\rangle=
  \left\langle
    \frac{1}{2}
    \left\{
      \left(
        \frac{d f_2}{dx}
      \right)^2+
      m_2^2 f_2^2(x)
    \right\}
  \right\rangle.
\end{equation}
Here we suppose that $f_1(x)=0$. Using Eq.~\eqref{enerdens} in
relativistic limit ($\omega\gg m_{1,2}$) we obtain that
\begin{equation*}
  \langle\rho_E\rangle=
  \frac{\omega}{L},
\end{equation*}
and we can identify $\omega$ with the energy of the system. Thus
the first term in Eq.~\eqref{Intens} is similar to the well known
formula for the transition probability in the two neutrino system.
It is interesting to mention that the second term in
Eq.~\eqref{Intens} contains the harmonic oscillations with the
frequency $(m_1^2+m_2^2)/4\omega$. Analogous additional term was
obtained in Refs.~\cite{BlaVit95,BlaHenVit99,BlaVit99,BlaPacTse03}
and was treated as the quantum field theory correction to the
Eq.~\eqref{TrProbPon}. However our approach demonstrates that this
term appears when one uses classical field theory. It results from
the accurate account of the Lorentz invariance.

Now let us discuss the case of the coupled fermion fields. The
Lagrangian for this system is expressed in the following way
\begin{equation*}
  \mathcal{L}(\bm{\nu})=
  \sum_{k=1}^{N}\mathcal{L}_0(\nu_k)+
  \left(
    \sum_{\substack{i,j=1 \\ i>k}}^{N}
    g_{ik}\bar{\nu}_i\nu_k+
    \text{h.c.}
  \right),
\end{equation*}
where $\bm{\nu}=(\nu_1,\dots,\nu_N)$, and
\begin{equation*}
  \mathcal{L}_0(\nu_k)=
  \bar{\nu}_k(i\gamma^\mu\partial_\mu-\mathfrak{m}_k)\nu_k.
\end{equation*}
We again should set the Cauchy problem for the system of
differential equations in question. However here one has to impose
only one initial condition since Dirac equation is the first-order
differential equation,
\begin{equation}\label{inicondDir}
  \nu_k(\mathbf{r},0)=\xi_k(\mathbf{r}).
\end{equation}
Analogously to the case of the scalar fields we can introduce the
mass eigenstates,
\begin{equation*}
  \nu_i(\mathbf{r},t)=
  \sum_{k=1}^{N}M_{ik}\psi_k(\mathbf{r},t).
\end{equation*}
The Lagrangian expressed in terms of the mass eigenstates has the
form,
\begin{equation*}
  \mathcal{L}(\bm{\psi})=
  \sum_{k=1}^{N}\mathcal{L}_0(\psi_k),
\end{equation*}
where
\begin{equation*}
  \mathcal{L}_0(\psi_k)=
  \bar{\psi}_k(i\gamma^\mu\partial_\mu-m_k)\psi_k,
\end{equation*}
is the Lagrangian for the field $\psi_k(\mathbf{r},t)$. Note that
masses $m_k$ differ from the masses of the fields $\nu_k$.

The solution of the Dirac equations for the mass eigenstates
fields can be expressed in the following way,
\begin{align}
  \psi_k(\mathbf{r},t) = &
  \int
  \frac{\mathrm{d}^3\mathbf{p}}{(2\pi)^{3/2}}
  \big[
    a_{s}(\mathbf{p})u_{s}(\mathbf{p})e^{-i\mathcal{E}_k t}
    \notag
    \\
    &
    \label{nusol}
    +
    b_{s}(-\mathbf{p})v_{s}(-\mathbf{p})e^{i\mathcal{E}_k t}
  \big]
  e^{i\mathbf{p}\mathbf{r}}.
\end{align}
Here $u_{s}(\mathbf{p})$ and $v_{s}(\mathbf{p})$ are the basis
spinors, $a_{s}(\mathbf{p})$ and $b_{s}(\mathbf{p})$ are the
indeterminate functions.

Now we should find the values of the $a_{s}(\mathbf{p})$ and
$b_{s}(\mathbf{p})$ $c$-number functions to satisfy the initial
condition given in Eq.~\eqref{inicondDir}. The calculations are
analogous to the previously discussed case of the coupled scalar
fields. Thus we arrive to the solution of the Dirac equations
which are valid for arbitrary functions $\xi_i(\mathbf{r})$,
\begin{align}
  \nu_j(\mathbf{r},t)= &
  \sum_{ik=1}^N
  M_{jk}
  \left(
  M^{-1}
  \right)_{ki}
  \notag
  \\
  &
  \label{nusolviaxi}
  \times
  \int \mathrm{d}^3\mathbf{r}'
  S_k(\mathbf{r}'-\mathbf{r},t)(-i\gamma^0)\xi_i(\mathbf{r}'),
\end{align}
where
\begin{equation*}
  S_k(\mathbf{r},t)=(i\gamma^\mu\partial_\mu+m_k)D_k(\mathbf{r},t),
  \quad
  x^\mu=(t,\mathbf{r}),
\end{equation*}
is the Pauli-Jordan function for a fermion field (see, e.g.,
Ref.~\cite{BogShi80p607}). In deriving of Eq.~\eqref{nusolviaxi}
we used the orthonormality conditions
\begin{gather*}
  u^{\dag}_s(\mathbf{p})u_r(\mathbf{p})=
  v^{\dag}_s(\mathbf{p})v_r(\mathbf{p})=\delta_{sr},
  \\
  u^{\dag}_s(\mathbf{p})v_r(-\mathbf{p})=
  v^{\dag}_s(\mathbf{p})u_r(-\mathbf{p})=0,
\end{gather*}
and the formulae for the summation over the spin indexes
\begin{align*}
  \sum_s u_s(\mathbf{k})u^{\dag}_s(\mathbf{k}) & =
  \frac{{\not p}+m}{2p^0}\gamma^0,
  \\
  \sum_s v_s(\mathbf{k})v^{\dag}_s(\mathbf{k}) & =
  \frac{{\not p}-m}{2p^0}\gamma^0.
\end{align*}
It is interesting to mention that the function $S_k(\mathbf{r},t)$
has the following property,
\begin{equation*}
  S_k(\mathbf{r},0)=i\gamma^0\delta^3(\mathbf{r}).
\end{equation*}
Thus the solution given in Eq.~\eqref{nusolviaxi} is consistent
with the initial condition \eqref{inicondDir}. Equation
\eqref{nusolviaxi} can be rewritten in the non-covariant form
which is, however, more convenient for the further analysis,
\begin{align}
  \nu_j(\mathbf{r},t)= &
  \sum_{ik=1}^N
  M_{jk}
  \left(
  M^{-1}
  \right)_{ki}
  \notag
  \\
  &
  \times
  \Big\{
    -\int\mathrm{d}^3\mathbf{r}'
    (\bm{\alpha}\bm{\nabla}_\mathbf{r})
    D_k(\mathbf{r}-\mathbf{r}',t)\xi_i(\mathbf{r}')
    \notag
    \\
    &+
    \int\mathrm{d}^3\mathbf{r}'
    \dot{D}_k(\mathbf{r}-\mathbf{r}',t)\xi_i(\mathbf{r}')
    \notag
    \\
    &
    \label{nusolviaxiNC}
    -
    i m_k\beta\int\mathrm{d}^3\mathbf{r}'
    D_k(\mathbf{r}-\mathbf{r}',t)\xi_i(\mathbf{r}')
  \Big\},
\end{align}
where we use common notations for the gamma matrixes,
$\bm{\alpha}=\gamma^0\bm{\gamma}$ and $\beta=\gamma^0$.

Just for simplicity we again discuss the case of two coupled Dirac
fields in the space with $1+1$ dimensions. Dirac equation in $1+1$
dimensional space was carefully studied in
Refs.~\cite{Bro63,Hag67,LowSwi71}. The gamma matrixes have the
form,
\begin{equation}\label{gamma1+1}
  \gamma^0=
  \begin{pmatrix}
    1 & 0 \\
    0 & -1
  \end{pmatrix},
  \quad
  \gamma^1=
  \begin{pmatrix}
    0 & 1 \\
    -1 & 0
  \end{pmatrix}.
\end{equation}

Now one should set the initial conditions. Let us assume that
$\xi_1(x)=0$ and $\xi_2(x)$ is expressed in the following way,
\begin{equation}\label{xi2}
  \xi_2(x)=
  \frac{1}{2}
  \begin{pmatrix}
    \cos(\omega x/2) \\
    \sin(\omega x/2)
  \end{pmatrix}.
\end{equation}
We obtained the expression for the evolution of the $\nu_1(x,t)$
which accounts for the exact dependencies on the particles masses.
However it appeared to be rather awkward. Nevertheless it can be
shown that the the third term in Eq.~\eqref{nusolviaxiNC} is
negligible in $1+1$ dimensional space. Indeed, let us consider,
for instance, the integral
\begin{equation}\label{Ixt}
  I(x,t)=\int_{-\infty}^{+\infty}
  \mathrm{d}y
  \thinspace
  D_k(x-y,t)
  \sin
  \left(
    \frac{\omega}{2}y
  \right).
\end{equation}
We remind that the Pauli-Jordan function in $1+1$ dimensional
space is given in Eq.~\eqref{Dk}. The integral in Eq.~\eqref{Ixt}
can be calculated explicitly and expressed in the form
\begin{equation}\label{IxtFin}
  I(x,t)=\sin
  \left(
    \frac{\omega}{2}x
  \right)
  \frac{\sin
  (t\sqrt{m_k^2+(\omega/2)^2})}
  {\sqrt{m_k^2+(\omega/2)^2}}.
\end{equation}
Here we used the known value of the integral
\begin{equation*}
  \int_0^a\mathrm{d}x
  \thinspace
  J_0(b\sqrt{a^2-x^2})\cos(cx)=
  \frac{\sin
  (a\sqrt{b^2+c^2})}
  {\sqrt{b^2+c^2}}.
\end{equation*}
Thus in the high energy approximation ($\omega\gg m_k$) we obtain
that $I(x,t)\to 0$. It is also interesting to note that one should
carefully follow the order of integration and differentiation
while using the Pauli-Jordan function. Indeed using, for example,
Eqs.~\eqref{Dkdot}-\eqref{IntermInt} and \eqref{IxtFin} we can see
that
\begin{equation*}
  \frac{d}{dt}
  \int_{-\infty}^{+\infty}
  \mathrm{d}x
  \thinspace
  D_k(x,t)f(x)
  \not=
  \int_{-\infty}^{+\infty}
  \mathrm{d}x
  \thinspace
  \frac{\partial}{\partial t}
  D_k(x,t)f(x),
\end{equation*}
because Pauli-Jordan function is the singular one.

Finally we get the expression for the $\nu_1(x,t)$,
\begin{align}
  \nu_1(x,t)= &
  \sin2\theta
  \left[
    \sin
    \left(
      \frac{\omega}{2}t
    \right)+
    \cos
    \left(
      \frac{\omega}{2}t
    \right)
  \right]
  \notag
  \\
  &\times
  \sin
  \left[
  \frac{t}{4\omega}(m_1^2-m_2^2)
  \right]
  \cos
  \left[
  \frac{t}{4\omega}(m_1^2+m_2^2)
  \right]
  \notag
  \\
  &
  \label{nu1}
  \times
  \begin{pmatrix}
    \cos(\omega x/2) \\
    \sin(\omega x/2)
  \end{pmatrix}.
\end{align}
In deriving of Eq.~\eqref{nu1} we used the fact that
\begin{equation}
  \frac{\partial}{\partial x}D_k(x,t)=
  -\frac{x}{t}
  \frac{\partial}{\partial t}D_k(x,t).
\end{equation}

The measurable quantity of the classical Dirac field is the
intensity. It is proportional to the $|\nu_1(x,t)|^2$. However we
again should average the intensity over space and time. Thus,
using Eq.~\eqref{nu1} we obtain for $\langle I\rangle(t)$ the
following expression
\begin{align}
  \langle I\rangle(t)= &
  \sin^2 2\theta
  \sin^2
  \left(
    \frac{\Delta m^2}{4\omega}t
  \right)
  \notag
  \\
  &
  \label{IntensDir}
  \times
  \left\{
    1-
    \sin^2
    \left(
      \frac{m_1^2+m_2^2}{4\omega}t
    \right)
  \right\},
\end{align}
which coincides with the similar expression derived for the scalar
field. Note that Eq.~\eqref{IntensDir} again contains the
additional term oscillating with the frequency
$(m_1^2+m_2^2)/4\omega$.

The calculations performed in this work demonstrate [especially
Eq.~\eqref{IntensDir}] that neutrino flavor oscillations can be
treated in frames of the classical theory. According to the
classical field theory approach the evolution of flavor neutrinos
is described in the following way.
\begin{enumerate}
  \item Flavor neutrino emission in a reaction.
  This process can be described by means of the quantum approach.
  However here we should obtain the final field distribution rather than
  the emission probability. It is also possible to admit that the
  mixture of the neutrino flavors appears in a process, as it was
  proposed in Ref.~\cite{Fie03}.
  In this case one should set other initial conditions in
  Eqs.~\eqref{inicond} and \eqref{inicondDir}.\label{emission}
  \item Neutrino propagation towards a detector. One can
  successfully use the methods elaborated in this paper for the
  description of the neutrino conversion or oscillations.
  Basing on the initial fields
  distributions obtained in the item~\ref{emission} we derive
  the final fields distributions that take into
  particles mixing.\label{oscillations}
  \item Neutrino interaction with a detector.
  This process again can be described by means of quantum field
  theory. On the basis of the fields distributions obtained in the
  item~\ref{oscillations} one can calculate the neutrino flux
  measured with a detector.
\end{enumerate}
Thus we should not directly involve the mass eigenstates if we are
using classical approach.

In conclusion we mention that the evolution of the coupled scalar
as well as fermion fields within the classical field theory has
been studied in this paper. We have examined the case of $N$
coupled fields in $1+3$ dimensional space. The Cauchy problem has
been formulated for these systems. We have solved it for arbitrary
initial conditions. The particular case of two coupled fields in
$1+1$ dimensional space has been studied. Finally we have obtained
the expressions for the averaged fields intensities. It has been
shown that in the relativistic limit these expressions were
similar to the usual transition probabilities formulae of neutrino
oscillations in vacuum. The discussion of the additional terms in
transition probabilities formulae has been presented. It has been
demonstrated that the expressions for the averaged fields
intensities for both bosons and fermions turned out to be
identical. We have shown by means of the direct calculations that
the flavor oscillations phenomenon could be described within the
classical approach. Thus one can conclude that the usage of the
quantum mechanics is inexpedient because classical field theory
yields more elegant description of the problem in question.

\begin{acknowledgments}
This research was supported by Deutscher Akademischer Austausch
Dienst. The author is indebted to Michael Beyer for making the
facilities of Universit\"{a}t Rostock available to him and for
hospitality.
\end{acknowledgments}

\bibliography{generaleng}

\end{document}